*Article*

# Elastic Characterization of Transparent and Opaque Films, Multilayers and Acoustic Resonators by Surface Brillouin Scattering: A Review

**Giovanni Carlotti**

Dipartimento di Fisica e Geologia, Università di Perugia, Perugia 06123, Italy; giovanni.carlotti@unipg.it; Tel.: +39-0755-852-767

**Abstract:** There is currently a renewed interest in the development of experimental methods to achieve the elastic characterization of thin films, multilayers and acoustic resonators operating in the GHz range of frequencies. The potentialities of surface Brillouin light scattering (surf-BLS) for this aim are reviewed in this paper, addressing the various situations that may occur for the different types of structures. In particular, the experimental methodology and the amount of information that can be obtained depending on the transparency or opacity of the film material, as well as on the ratio between the film thickness and the light wavelength, are discussed. A generalization to the case of multilayered samples is also provided, together with an outlook on the capability of the recently developed micro-focused scanning version of the surf-BLS technique, which opens new opportunities for the imaging of the spatial profile of the acoustic field in acoustic resonators and in artificially patterned metamaterials, such as phononic crystals.

## 1. Introduction

The problem of a detailed knowledge of the elastic constants of thin film materials has attracted much attention in recent years, because of the growing importance of single- and multi-layered structures in advanced applications such as in devices for information and communication technology (ICT). For instance, current filters and duplexers in mobile communication devices exploit either bulk or surface acoustic wave (BAW or SAW) resonators operating at several GHz [1–4], whose design and optimization require a detailed knowledge of the elastic properties of the constituent layers. In fact, as illustrated in Figure 1, free-standing bulk acoustic wave resonators (FBAR) and solidly mounted resonators (SMR) consist of a stack of several materials, including the active resonator (usually a film of AlN, ScAlN or ZnO), metallic layers for electrical electrodes (Ti, Au, etc.) and other materials for temperature compensation ($SiO_2$) or for Bragg mirrors ($SiO_2$ and W). The thicknesses of the above layers range from a few tens of nanometers up to a few microns. It follows that it is crucial to dispose of advanced techniques and methods to provide information about the elastic constants of the constituent layered materials and to image the spatial profile of the acoustic field, both inside and outside the resonator.

In this paper, I will review the capability of the surface Brillouin light scattering technique (surf-BLS) to the above aims and in particular to successfully achieve a selective determination of the elastic constants of various types of films and multilayers. This technique is based on the inelastic scattering of photons by thermal phonons, naturally present in any material at room temperature, through the modulation of the optical constants of the medium, which is induced in transparent media via the elasto-optic effect [5,6]. However, in the case of opaque media, either metals or semiconductors, the rippling of the free surface, caused by the presence of thermal phonons, gives a substantial contribution to the light scattering [7]. In fact, the periodic modulation of the optical constant (either in the interior or at the surface of a medium) associated to a propagating acoustic wave results in a moving grating so that Brillouin scattering can be understood in very simple terms as a Doppler shift of the diffracted light. From the quanto-mechanical point of view, however, one considers a photon-phonon collision in which the frequency and the momentum are conserved through phonon creation or annihilation, corresponding to either Stokes or anti-Stokes processes, respectively. It follows that in a typical Brillouin light scattering experiment, the probed phonons have wavelength comparable to that of light sand this implies that their frequencies are in the range between about 10 and 150 GHz. The cross section of the scattering process is relatively low, so that in order to extract the tiny inelastic component of light from the elastically scattered contribution, a high-resolution spectrometer is required. To this aim, the best combination of high resolution and good throughput is achieved using a multipass Fabry-Perot interferometer.

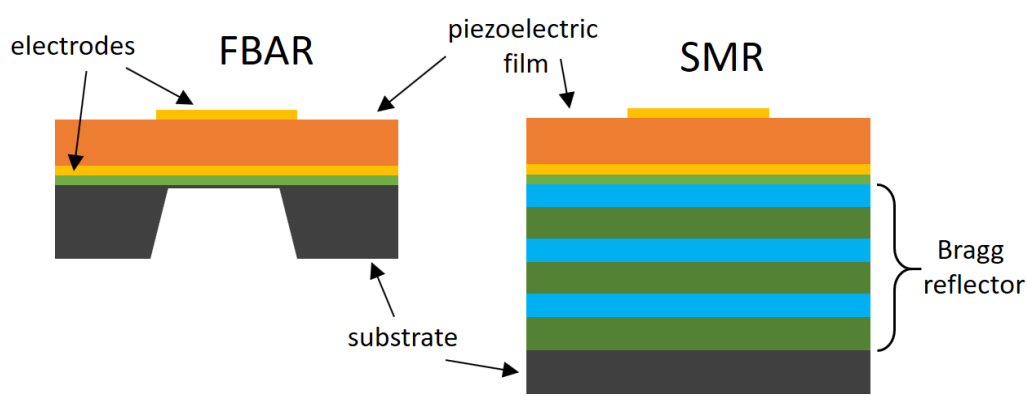

**Figure 1.** Schematic drawings of free-standing (left) and solidly mounted (right) bulk acoustic-wave resonators, named free-standing bulk acoustic wave resonators (FBAR) and solidly mounted resonators (SMR), respectively, that are currently used in filters and duplexers for mobile communication devices.

Measurement of the whole set of constants is usually out of the reach of conventional static techniques, such as micromechanical test or nanoindentation, which can only give information about specific elastic moduli [8]. On the other hand, use of acoustic techniques based on surface acoustic waves (SAW) presents technical difficulties connected with the fabrication of the acoustical delay line. In addition, one has to consider that in the usual frequency range of hundreds of MHz, the large values of the acoustic wavelength, compared to the film thickness, require a careful consideration of the effect of the substrate on the propagation of the SAW [9]. These limitations can be overcome using surf-BLS, because it has specific advantages over conventional techniques. In particular,

- it does not require any external generation of acoustic waves (since it probes thermal acoustic phonons naturally present into the medium);
- it is a local probe, since the investigated portion of the sample is defined by the dimension of the focused laser spot and this permits to operate a scan of the elastic properties of inhomogeneous specimens;
- it is suited to analyze short wavelength and high-frequency phonons, up to several GHz, such as those of current and forthcoming ICT devices operating at microwave frequencies.

The latter characteristic descends from the fact that, thanks to the wavevector conservation in the photon-phonon interaction, the wavelength of the revealed elastic waves is of the same order of magnitude of that of light. It is interesting to notice that most of the experimental investigations of layered structures, such as supported films and superlattices, have been carried out in the eighties and early nineties, i.e., just after the availability of the high-contrast Tandem Fabry-Perot Interferometer. Later on, around the turning of the millennium, thanks to the growing importance of the rising fields of spintronics and magnonics, most of the research groups that had been active in the application of surf-BLS to acoustic phonons [such as the group of B. Hillebrands in Kaiserslautern (Germany), that of M. Grimsditch in Argonne (USA), and the groups of F. Nizzoli in Ferrara and mine in Perugia (Italy)] switched their main interest to the study of magnetic films and multilayers by detection of spin waves. In some sense, this caused a slowing down in the transfer of expertise and in the education of young researchers in the application of surf-BLS to the elastic characterization of layered samples. The result is that, according to my experience, there is currently a gap between the needs of the industrial manufacturers of advanced acoustic components (such as filters and duplexers) for mobile telecommunication devices and the availability of experts in the application of surf-BLS to their needs. The main aim of this paper is just to provide a contribution to fill this gap.

The structure of the paper is as follows. In Section 2, I recall the physical mechanism that is at the origin of the experiments, focusing on the interaction between light and acoustic phonons and discussing in details the quantitative relations that hold in the backscattering geometry that is the most popular for the study of opaque media. Section 3 is devoted to the presentation of the situations that may occur when applying the surf-BLS technique to the elastic characterization of different types of films. In particular, the experimental methodology that apply to specific case studies are discussed together with the discussion of the amount of information that can be extracted, depending on the thickness of the film and on its transparency. A short presentation of the limitations of the technique and a generalization to the case of multilayered samples is also provided. In the last Section, the recent development of the micro-focused, scanning, version of the surf-BLS technique are shortly discussed, addressing the new opportunities that are at hands for the imaging of the spatial profile of the acoustic field in acoustic resonators and phononic crystals, with a deeply submicrometric resolution.

## 2. Brillouin Scattering from Thin Films and Opaque Media: Historical Background and Interaction Mechanism

As introduced in the previous paragraph, the surf-BLS technique is presently the most powerful tool to access the elastic characteristics of thin films and layered structures in the GHz range of is frequencies that are of growing importance in nowadays ICT devices. The physical principle that stands behind this technique was initially argued, almost a hundred years ago, by the pioneering and independent works of Leon Brillouin and Leonid Mandelstam [10,11] concerning the scattering of light by acoustic waves in liquids and transparent media. Extension to thin films and metallic or semiconductor solids (i.e., opaque to visible light) were achieved in the seventies thanks to both the advent of lasers and the construction of high-contrast Fabry-Perot interferometers [12–14]. John R. Sandercock developed an original type of these machines in the so-called "tandem" configuration during the seventies, while working in the research division of the RCA company in Zurich, but a few years later he established his own company (formerly named "JRS Scientific Instruments" and now "The Table Stable") [15], which launched into the market the multi-pass tandem Fabry-Perot interferometer (TFP). After more than thirty years from its commercialization, this kind of interferometer is still without alternatives on the market for the detection of either phonons or magnons from thin films and opaque solids. Several upgrades of the setup have been performed during the last two decades to achieve better performance in terms of contrast, resolution and signal-to-noise ratio.

Let us now recall some relevant characteristics of the physical mechanism lying below the surf-BLS from acoustic waves. As stated in the Introduction, it is based on the coupling between the electric field of a monochromatic incoming light beam and the periodic variation of the dielectric constant of the medium associated with acoustic waves. The modulation of the dielectric constant of the medium, induced by the acoustic wave, can be described according to the following equation:

$$\Delta\varepsilon_{\alpha\beta} = k_{\alpha\beta\gamma\delta}\, S_{\gamma\delta} \qquad (1)$$

where $S_{\gamma\delta}$ is the strain field associated with the wave and $k_{\alpha\beta\gamma\delta}$ is the elastooptic tensor. Following the derivation in Ref. [16], the scattered electric field *Es* at large distance from the scattering volume *V* is proportional to the following integral:

$$\iiint_V \Delta\varepsilon_{\alpha\beta}\, \boldsymbol{E}_i\, e^{-i(\boldsymbol{K}_i - \boldsymbol{K}_s)\cdot \boldsymbol{r}}\, dV \qquad (2)$$

where $E_i$ is the electric field of the incident light, $K_i$ and $K_s$ are the wavevectors of the incident and scattered light. The above integral leads directly to the conservation of momentum, because for a given acoustic excitation of wavevector $Q$, $\Delta\varepsilon_{\alpha\beta}$ is proportional to $e^{-iQ\cdot r}$ so that, if the interaction volume $V$ is large with respect to the acoustic wavelength, the integral becomes a delta function $\delta(K_i - K_s \mp Q)$ and this means that the exchanged wavevector must be conserved.

As for the intensity of the inelastic scattering from a bulk acoustic wave in an isotropic medium at temperature $T$, the scattering cross section (scattered power divided by the solid angle), for an unitary interaction volume and incident power, can be expressed as [17]:

$$\frac{dP}{d\Omega} = \frac{k_B T \pi^2}{4\varrho v^2 \lambda^4} |g|^2 \qquad (3)$$

where $\varrho$ is the mass density, $\lambda$ is the light wavelength and $v$ the acoustic velocity. Interestingly, $g$ is a vector that is determined by the values of the contracted elastooptic coefficients $k_{\alpha\beta}$ [18] according to:

$$g_\alpha = \frac{1}{QE_s}\left[(k_{11} - k_{12})(e_\alpha Q \cdot E_i + Q_\alpha e \cdot E_i) + 2k_{12}\delta_{\alpha\beta}E_\beta Q \cdot e\right] \qquad (4)$$

where $e$ is the unit polarization vector of the acoustic wave. It results that the scattered intensity depends on the values of the elastooptic coefficients and on the polarization direction of the incoming and scattered light beams, as well as of the acoustic wave. However, one can derive the following simple guiding rule. Longitudinal acoustic waves give origin only to polarized scattering, i.e., the light polarization of the scattered photons is the same of that of incoming photons. In particular, if the incoming and scattered beams are both polarized perpendicular to the plane of incidence, s-s scattering, one has that the cross section is proportional to $|g|^2 = 4k_{12}^2$, while for p-p scattering one has $|g|^2 = 4\left|(k_{11} - k_{12})\cos^2\frac{\psi}{2} + k_{12}\cos\psi\right|^2$, where $\psi$ is the scattering angle, formed by $K_i$ and $K_s$.

Instead, shear horizontal (transverse) waves that are polarized perpendicular to the scattering plane, give origin only to depolarized scattering (p-s or s-p scattering), and the cross section is proportional to $|g|^2 = (k_{11} - k_{12})^2 \cos\frac{\psi}{2}$.

From the brief description above, it follows that the polarization of the light scattered by acoustic waves can be either parallel or perpendicular to that of the incident light and this fact is of great help for the experimentalist in order to select the scattering from specific acoustic modes, as we will show in the following. Moreover, as anticipated in the Introduction, there is another interaction mechanism that contributes to the light scattering and is particularly relevant in opaque samples: the rippling of the free surface, or of the interface of a thin film, induced by the acoustic phonons. It contributes to polarized scattering, only, but the relative importance of the elastooptic and the ripple mechanism depends on the specific medium under investigation, with a prevalence of the former (latter) in the case of transparent (opaque) media, as we will see in the following [6,7].

From the experimental point of view, the most simple and popular surf-BLS interaction geometry, suitable for thin films and opaque samples, is that represented in of Figure 2, called backscattering geometry, where the scattering angle $\psi = \pi$ and the same lens (typically a camera objective with f number comprised between 2 and 4 and focal length of 35 or 50 mm) is used for focusing the laser beam on the sample and for collecting the light that is back-scattered within a solid angle. A discussion of the influence of the *f* number of the collecting lens on the measurement accuracy can be found in Refs. [19,20]. Two classes of phonons contribute to the scattering in this geometry. First, through the conservation of the momentum component parallel to the surface, one can reveal acoustic waves propagating parallel to the free surface ("surface-like" waves) with an acoustic wavenumber $Q_S = 2k_i \sin(\theta)$ with $k_i = 2\pi/\lambda$, with $\lambda$ the optical wavelength (typically 514.5 or 532 nm) and $\theta$ the angle of incidence of light. Moreover, when the film thickness is much larger than the wavelength of light, "bulk-like" acoustic waves with wavevector $Q_B = 2nk_i$ enter the scattering process, with n the refractive index of the film. Due to optical refraction, these bulk waves propagate within the film material at an angle $\alpha = \sin^{-1}(\sin\theta/n)$, as seen in Figure 2.

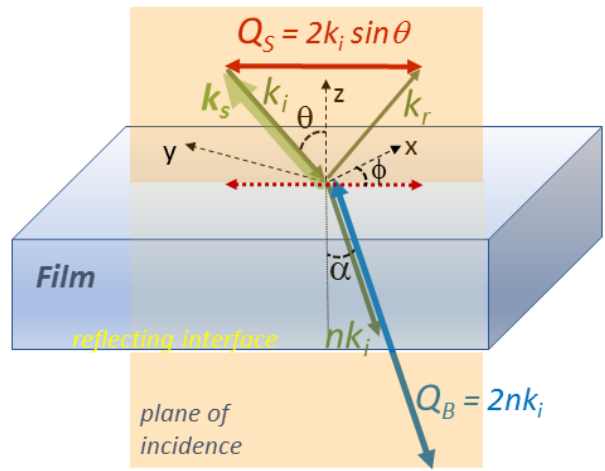

**Figure 2.** Back-scattering geometry for conventional surface Brillouin light scattering (surf-BLS) experiments. The wave vectors of the incident, reflected, transmitted and scattered light are indicated in green, while those of acoustic waves are in red (surface waves) or in blue (bulk waves). All are contained in the plane of incidence.

One can therefore directly derive the connection between the frequency shift f of Brillouin peaks and the phase velocity v of the corresponding acoustic mode, as follows:

$$v_S = 2\pi f/Q_S = \lambda f/2\sin(\theta) \tag{5}$$

for surface-like modes, and

$$v_B = 2\pi f/Q_B = \lambda f/2n \tag{6}$$

for bulk-like modes.

### 3. Determination of the Elastic Constants of Nanometric or Micrometric Films

Let us now consider the case of a film with either micrometric or submicrometric thickness (such as those used in acoustic-wave resonators), whose elastic constants are to be determined. We will see in the following that the kind of acoustic modes that can be detected in a surf-BLS experiment, and the consequent amount of information that can be achieved, depends on the thickness of the film, its transparency and the presence of a substrate where the acoustic velocity is larger or smaller than in the film. It follows that the above conditions must be carefully considered if one targets a complete elastic characterization of thin films in terms of the whole set of independent elastic constants. Note that even in the case of opaque media, where the penetration depth of light can be as small as ten nanometers, the gained information concerns the whole coherence depth of the acoustic waves, which is of the order of the light wavelength, i.e., hundreds of nanometers. Therefore, for submicrometric film thickness the influence of substrate on the measured frequencies is substantial and must be properly modeled if one wants to extract information about the films. This is different from the case of spin waves in ferromagnetic films, because in that case the presence of a non-magnetic substrate is completely non-influent [21].

*3.1. Transparent Films with Micrometric Thickness: Guided Modes and Selective Determination of the Elastic Constants*

The most favorable situation occurs when one deals with a single transparent film of micrometric thickness, supported by a reflecting substrate (such as optically polished Si or GaAs wafers). In this case, it is possible to obtain, in a relatively straightforward way, the whole set of independent effective elastic constants. These are only two for acoustically isotropic films, such as amorphous glasses and soft dielectric films used in electronics, while may increase to three independent constants in the case of epitaxial cubic films, or to five for films of hexagonal (cylindrical) symmetry, such as piezoelectric films with a textured polycrystalline structure.

Let us now consider the latter (more complex) case that has been considered in previous investigations devoted, for instance, to ZnO [22] or AlN [23,24] films. Using different angles of incidence and a careful polarization analysis of the scattered light, one can detect and identify several guided acoustic modes [25], such as the surface Rayleigh wave (RW), the shear horizontal mode (SHM), the longitudinal mode (LM) and the longitudinal bulk (LB) mode, whose propagation direction and polarization are summarized in Figure 3. (Please note that in some papers the LM is labeled as LGM, longitudinal guided mode, and the LB is labeled as LA, longitudinal acoustic wave). Their detection provides a sufficiently large amount of information to estimate the whole set of independent elastic constants, because as summarized in Figure 3, measuring the frequency of the Brillouin peaks associated to the above mentioned modes provides access to the values of the five independent constants, namely $C_{11}$, $C_{13}$, $C_{33}$, $C_{44}$ and $C_{66}$. In particular, one has the following selective relationship between the measured velocities and the elastic constants: $V_{LM} = \sqrt{\frac{C_{11}}{\rho}}$, $V_{SHM} = \sqrt{\frac{C_{66}}{\rho}}$ and $V_{RW} = \beta \sqrt{\frac{C_{44}}{\rho}}$ with $\beta$, a constant that depends on the other elastic constants $C_{11}$, $C_{13}$, $C_{33}$. It must be evaluated numerically, using the model of [25] and usually takes values in the range between about 0.9 and 0.95. The exact value of $\beta$ can be found according to the following implicit expression derived by Dobrzynski and Maradudin [26]:

$$C_{33}\left(v_R^2 - \frac{C_{44}}{\rho}\right)\left(v_R^2 - \frac{C_{11}}{\rho} + \frac{C_{13}^2}{\rho C_{33}}\right) = C_{44} v_R^4 \left(v_R^2 - \frac{C_{11}}{\rho}\right) \qquad (7)$$

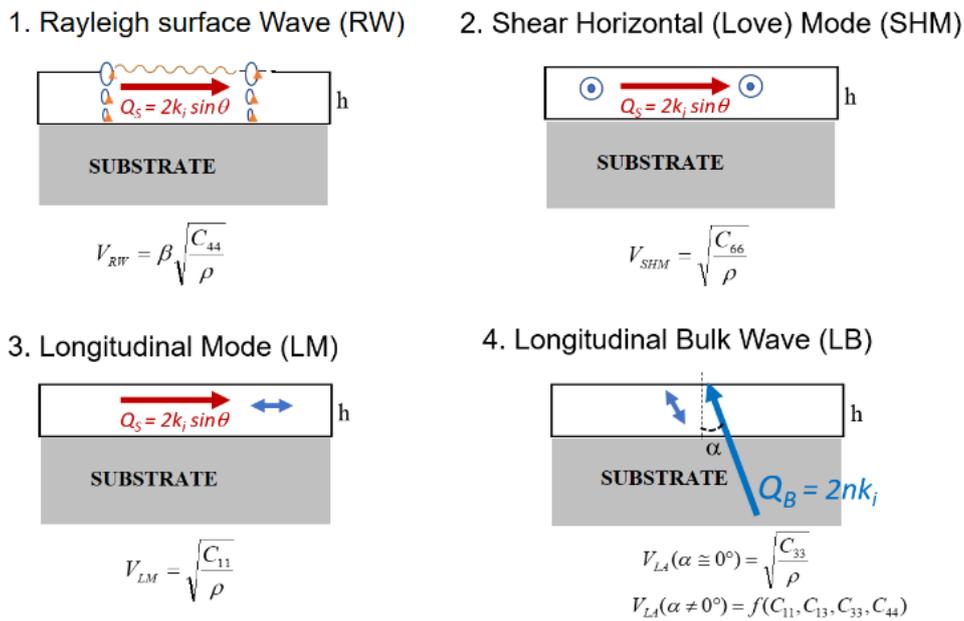

**Figure 3.** Schematic view of the four kind of acoustic modes that can be detected in Brillouin spectra taken from transparent films of micrometric thickness deposited on a reflecting substrate. The Rayleigh wave (RW), the longitudinal mode (LM) and the longitudinal bulk (LB) waves are polarized in the sagittal plane, while the shear horizontal mode (SHM) is polarized perpendicular to the sagittal plane, so that the latter gives origin to depolarized light scattering.

To illustrate the above considerations, in Figure 4, one can see typical spectra relative to an AlN film of 1.7 μm thickness measured on a wide frequency range, up to 100 GHz, where several peaks are present. Some of them, corresponding to the Rayleigh wave (RW) and the longitudinal mode (LM), change their frequency position with θ, according to Equation (5). The peak corresponding to the longitudinal bulk wave (LB) instead, stays almost fixed with changing θ, because it corresponds to the interaction mechanism of bulk phonons, according to Equation (6). A deeper insight into the characteristics of "surface" waves show that the peak at lower frequency is actually a doublet, corresponding to scattering from either the Rayleigh wave (polarized scattering, p-p) or the shear horizontal mode (SHM) that is depolarized (p-s) [22]. Measuring the frequency shift f of each of these two peaks, one can determine the phase velocity of the corresponding acoustic modes according to Equation (5). For the LB peak, instead, the corresponding phase velocity is obtained by Equation (6). Measuring the LB wave at normal incidence permits to obtain $C_{33} = \varrho \times V_{LB}^2$. Finally, the value of $C_{13}$ can be estimated, even if less precisely, from the dependence of $V_{LB}$ on the angle α, performing measurements at different incidence angles. Note however that if one uses the above procedure, disregarding the contribution of the piezoelectric constants, one obtains the values of the so called "piezoelectrically-stiffened" effective elastic constants. If one instead wants to estimate the "true" elastic constants values, one should also consider the contribution of the piezoelectric terms in the numerical model and this implies that the values of $C_{11}$, $C_{13}$, $C_{33}$ are slightly modified [24].

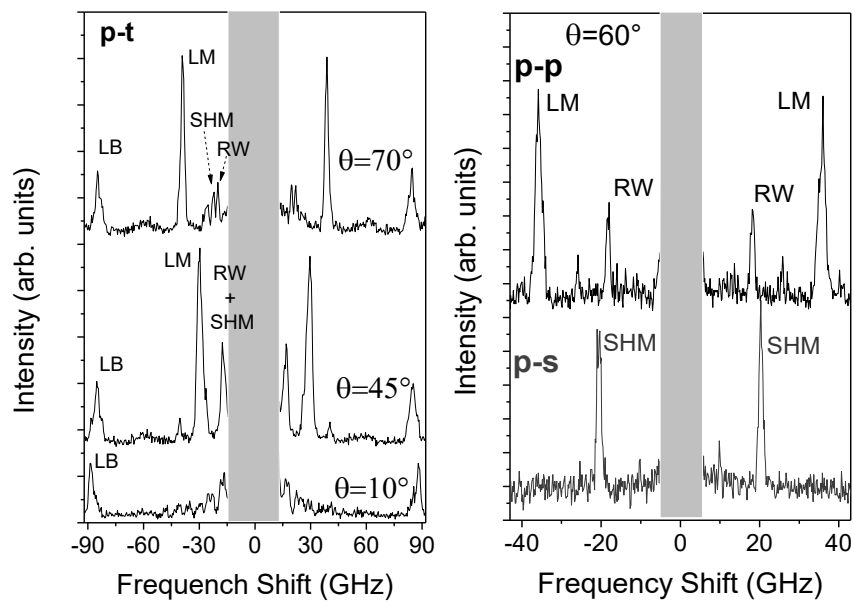

**Figure 4.** (Left) Comparison between two Brillouin spectra, relative to a 1.7 μm thick AlN film, supported by Si, extending to a large frequency range, collected at different angles of incidence. The labels LM and LB refer to the longitudinal mode and the longitudinal bulk acoustic waves, respectively. The incoming light was to p-polarized and no polarization analysis was made on the scattered light; (Right) Comparison between two Brillouin spectra, corresponding to p-polarized incoming light and p- or s-polarized scattered light (labeled p-p and p-s, respectively). It is seen that in the former case one can detect the peaks corresponding to the surface Rayleigh wave (RW) and to the longitudinal mode (LM), while in the latter case, only the peak relative to shear horizontal mode (SHM) is detected. The angle of incidence of light was ϑ = 60°. Adapted with permission, from [24] IEEE, 2017.

Please note that if the studied film has only two [27,28], or three [29] independent elastic constants, it is not necessary to look for the depolarized scattering from the SHM (that is much less intense than the other modes [30]), because the information extracted from the RW and either LM or LB is sufficient to determine the two independent constants $C_{11}$ and $C_{44}$.

In any case, even if the polarization analysis of the detected light is not necessary, one should notice that the cross section of the RW and LM can be rather strongly dependent on both the value of the incidence angle and the polarization status of the incoming light. The RW whose cross section mainly relies on the rippling mechanism, can be efficiently detected using an incoming p-polarization and a relatively large angle of incidence (θ ≥ 50°) [31]. Instead, the cross section of the LM depends on the values of the photoelastic constants of the material under consideration. For instance, the cross section of the LM in dielectric glasses is much stronger using s-polarization (rather than p-polarization) of the incoming light at relatively small angles of incidence (30° < θ < 50°). This is clearly visible in the spectra reported in Figure 5, taken from Ref. [27].

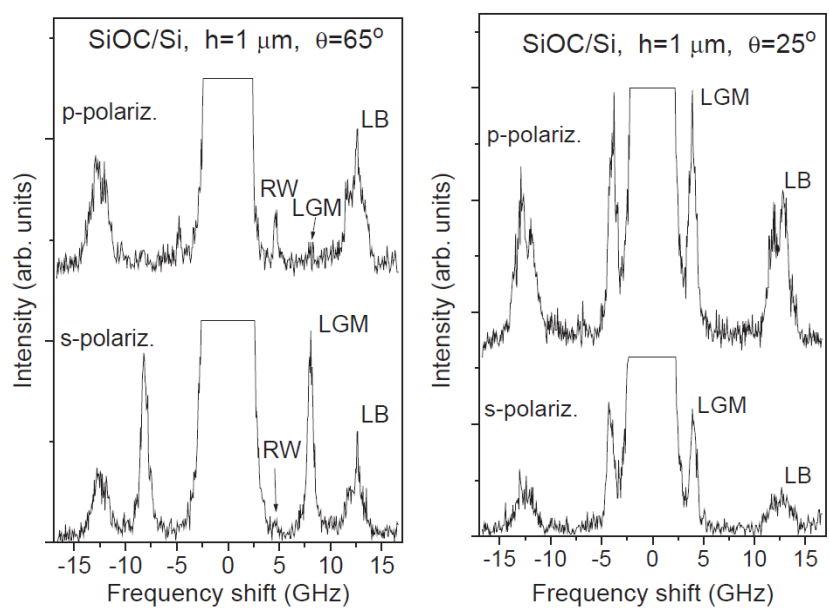

**Figure 5.** Experimental surf-BLS spectra relative to the SiOC:H/Si film, taken for an angle of incidence of light θ = 65° (left panel) and 25° (right panel). The upper spectra were taken using p-polarized incident light, while s-polarized light was used for the lower ones. The peaks corresponding to the Rayleigh surface wave (RW), to the longitudinal guided mode (LGM) and to longitudinal bulk wave (LB) are indicated on the anti-Stokes side of each spectrum. It can be seen that the RW peak has a large cross section for p-polarized incoming light and large angle of incidence. Instead, the LGM has a larger cross section for s-polarized light and small angle of incidence. Reprinted with permission from [27], Elsevier, 2005.

*3.2. Transparent Films with Deeply Submicrometric Thickness: Dispersion Curves and Interference Effects*

A relatively different situation occurs when the film thickness is submicrometric, i.e., comparable to, (or lower than) the acoustic wavelength. In the usual case of a film supported by a substrate, whose acoustic phase velocities are higher than those of the film (slow film/fast substrate), a number of discrete acoustic modes, namely the Rayleigh-Sezawa and Love modes are present [25] and can be revealed in Brillouin spectra. The former modes are polarized in the sagittal plane (i.e., in the plane of incidence of light, while the latter modes are polarized perpendicularly to the above plane). This transformation from a spectrum typical of thick films to the discretized spectrum typical of thin films is seen in the left panel of Figure 6, relative to GaSe films of different thickness [32]. Here, it is seen that the Rayleigh and a few Sezawa discrete modes are seen for film thickness of 77 and 251 nm. Taking measurements at different angles of incidence, one can obtain the dispersion curves shown in Figure 6, right panel.

A detailed theory about the theoretical cross section of these modes has been presented in [33], and it has been shown that strong interference effects can arise when one considers all the possible scattering mechanisms (ripple scattering from both the free surface and the interface, elastooptic scattering from the interior of the film) [34]. However, these modes are dispersive, so that measurements are usually performed on films of different thicknesses and with different angles of incidence, in order to span a large interval of the ratio between the film thickness and the phonon wavelength. In the case of a hexagonal elastic symmetry, four of the five effective elastic constants, namely $C_{11}$, $C_{13}$, $C_{33}$ and $C_{44}$, influence the Rayleigh-Sezawa modes, so that they can be evaluated by a best fit procedure of the experimental velocities to the calculated dispersion curves. This procedure has been used in the past for a number of different thin film materials, such as cubic Zinc Selenide [35], hexagonal and cubic

Boron Nitride [36–38], hexagonal Zinc Oxyde [39], Tin Dioxide [40], Gallium- [32] or Indium- [41] Selenide, and isotropic dielectrics [42–44]. In all the above studies it has been put in evidence that the different elastic constants can influence the acoustic modes in a similar way, so that one has to face the problem of the correlation among the fitting parameters. It follows that is useful to exploit films of different thickness (in the range between a few tens to a few hundreds of nanometers), but in his case a further complication may arise, if the structural and elastic differences among films of different thickness are not negligible, due for instance to the occurrence of a modified transition layer at the interface between substrate and film material. This is illustrated in Figure 7, where it is shown that the experimental data for the ZnO/Si films fit the expected theoretical dispersion curves fairly well for film thicknesses greater than 150 nm, while they appreciably depart from the same curves for smaller thicknesses [39]. Such a behavior (that was present in the case of Si substrate but not for a fused quartz one) was interpreted in terms of a reduction of the effective elastic constants of the film in a layer near the interface, due to the lattice misfit between the film and the Si substrate.

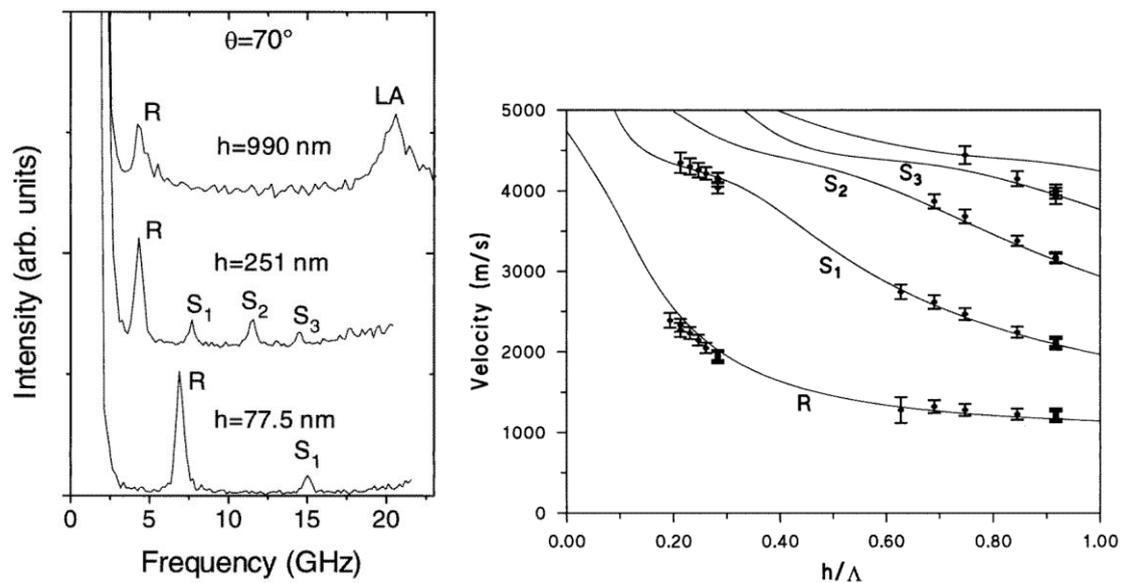

**Figure 6.** (Left) Brillouin spectra relative to the three GaSe films of different thickness *h*, for an angle of incidence θ = 70°. The peaks corresponding to the Rayleigh (R) and Sezawa (S) guided acoustic modes are clearly seen. In the case of the thicker film, h = 990 nm, the discrete Sezawa modes are not resolved anymore and the peak due to bulk longitudinal acoustic waves (LA) is observed; (Right) Experimental values of the phase velocity of the Rayleigh and Sezawa acoustic modes (points with error bars) as a function of the ratio between the film thickness and the acoustic wavelength. The dispersion curves (solid lines) have been obtained from a best fit procedure assuming the elastic constants of the film material as free parameters. Adapted with permission from [32], IOP Publishing, 1999.

In the above examples of Figures 6 and 7, a best fit procedure of the measured dispersion curves to the calculated ones permitted to achieve a determination of four of the five independent elastic constants. As for thick films discussed in the previous section, the fifth elastic constant $C_{66}$ can be determined from measurement of the phase velocity of shear horizontal modes (Love modes) in a depolarized scattering experiment, as shown in Figure 8 for an InSe film [41]. Use of an opaque substrate (typically crystalline silicon) is recommended in order to take advantage of the presence of a reflecting interface which enhances the Brillouin cross section, even if long acquisition times (several hours) are usually necessary.

To conclude this Section, let us note that in the case of a soft film on a hard substrate [25], instead, one observes only the RW, which evolves into a leaky wave as soon as the film thickness exceeds a few nanometers, radiating energy into the bulk (as for AlN on Si [45], CaF2 on GaAs [46] or TiN over high-speed steel [47]). As a consequence, the RW peak broadens and finally disappears from the spectra. In these cases, only a relatively limited information about the shear elastic constant $C_{44}$ can be extracted from the measurement.

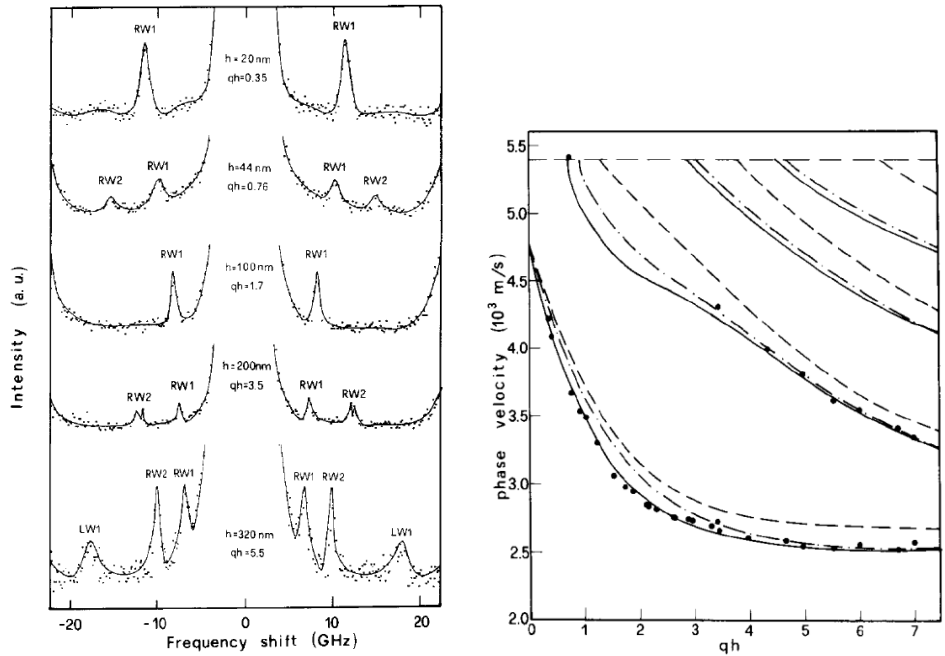

**Figure 7.** (Left) Brillouin spectra from ZnO films of different thickness on (001)-cut Si substrate for fixed backscattering angle θ = 45°; (Right) Experimental data of phase velocity of first two Rayleigh modes for the ZnO/Si structure and theoretical dispersion curves, plotted as a function of the product between the acoustic wavevector and the film thickness, evaluated by use of elastic constants of bulk ZnO (dashed lines), of effective constants of the film (dash-dotted lines) and of modified constants of the film (solid line). Adapted with permission from [39], IEEE, 1991.

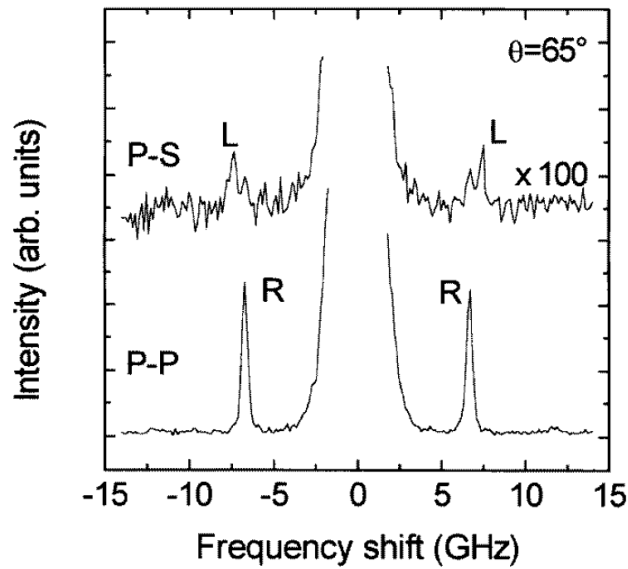

**Figure 8.** Brillouin spectra taken at an angle of incidence θ = 65°. From an InSe, 79 nm thick, film. The peaks corresponding to the Rayleigh wave (R) and to the Love mode (L) are clearly seen in the p-p spectrum and the p-s spectrum, respectively. Note that the ghost of the Rayleigh peak is also present in the p-s spectrum, because of the finite extinction ratio of the polarization analyzer. Reprinted with permission from [41], IOP Publishing, 1999.

*3.3. Opaque Films: Rippling of the Surface, Guided Modes and Relevant Thresholds*

The surf-BLS technique can be used also to determine the elastic constants of metallic films, where the penetration depth is only a few nanometers, so that the ripple mechanism is the dominant one while the elastooptic coupling is usually negligible. It is, therefore, recommendable to use p-polarized incoming light and relatively large angles of incidence to maximize the cross section [31].

The first investigations of opaque films were concerned with Au films of thickness from ten to hundreds of nanometers, supported by glass [48] where the Rayleigh and Sezawa guided modes were detected and the elastic constants estimated from a best fit procedure of their dispersion curves. In a successive study, the surf-BLS cross section of Au films on Si [49] was also analyzed, showing that the ripple mechanism gives the main contribution to the scattering process, even if the elastooptic contribution is not insignificant. Further studies of metallic films dealt with Al/NaCl [50] (where the authors were unable to fit our experimental results to obtain the elastic constants independently, since different combinations of the Cij lead to similar spectra) and with Mo [51] or Ni [52] films. More recently, other investigation on the limits and accuracy of the estimation of the constants have been proposed, with reference also the uncertainty connected with the film thickness [38,53,54].

Particularly relevant, during the late eighties and the early nineties, were the investigations of metallic superlattices, i.e., multilayers with many repetitions of elemental nanometric bilayers, where anomalous values of the elastic moduli were reported by conventional methods. Among the systems successfully investigated we recall here periodic superlattices: Mo/Ta [55], Cu/Nb [56] (quasi-periodic [57]), Ag/Pd [58], Fe/Pd [59], Co./Cu [60]. Co./Au [61], Ag/Ni [62], FeNi/Cu and FeNi/Nb [63] and Ta/Al [64] (quasi-periodic [65]). In all these systems, the overall film thickness was typically hundreds of nanometers, while the artificial periodicity ranged from a few nanometers to a few tens of nanometers. Therefore, the superlattice was modeled as an effective medium with hexagonal symmetry and five effective elastic constants, whose dependence on the artificial periodicity was investigated. Some of these constants, such as the shear constant $C_{44}$, was found to be appreciably dependent of the artificial periodicity, because of strong interface effects. The analysis was generally based on the analysis of the dispersion curves of Rayleigh and Sezawa modes, as shown in Figure 9 for the case of Ag/Ni superlattices. Four of the five constants elastic constants ($C_{11}$, $C_{13}$, $C_{33}$ and $C_{44}$) could be obtained by a best fit procedure of the measured frequencies on the calculated curves, while $C_{66}$ was not accessible since it only affects Love modes, not detected in the case of metallic samples.

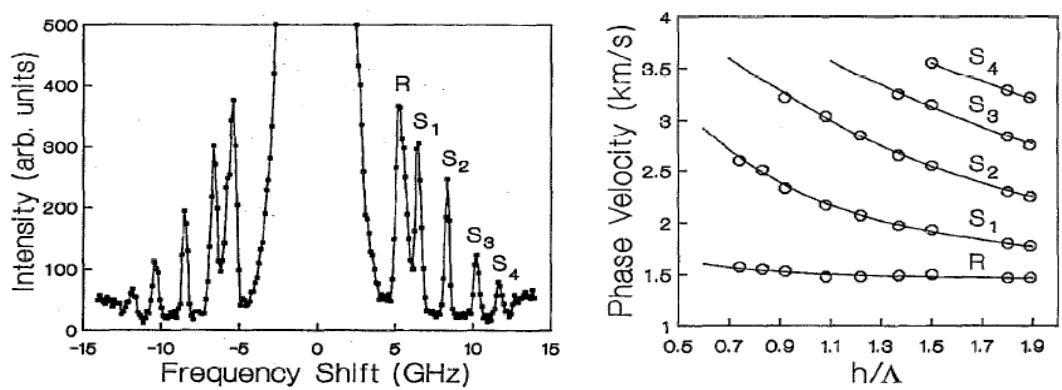

**Figure 9.** (Left) Brillouin spectrum taken on an Ag/Ni superlattice, 500 nm thick, having a periodicity of 2 nm, supported by a glass substrate. The Rayleigh and four Sezawa modes are seen. The incidence angle is 67°, which corresponds to a value of h/Λ = 1.8; (Right) Experimental (circles) and calculated (solid line) velocity dispersion of the Rayleigh and Sezawa modes for the same sample. Adapted with permission from [62], AIP Publishing, 1992.

In Figure 10 it is shown the effect of the different constants on the different modes, so one can see for instance that the shear constant $C_{44}$ gives the main contribution to the velocity of the Rayleigh mode at large film thickness, while its contribution to high-order modes is rather limited. One can also see that the contribution of $C_{11}$ and $C_{33}$ to the various modes is similar, so that it is not straightforward to disentangle these two constants from each other in the fitting procedure.

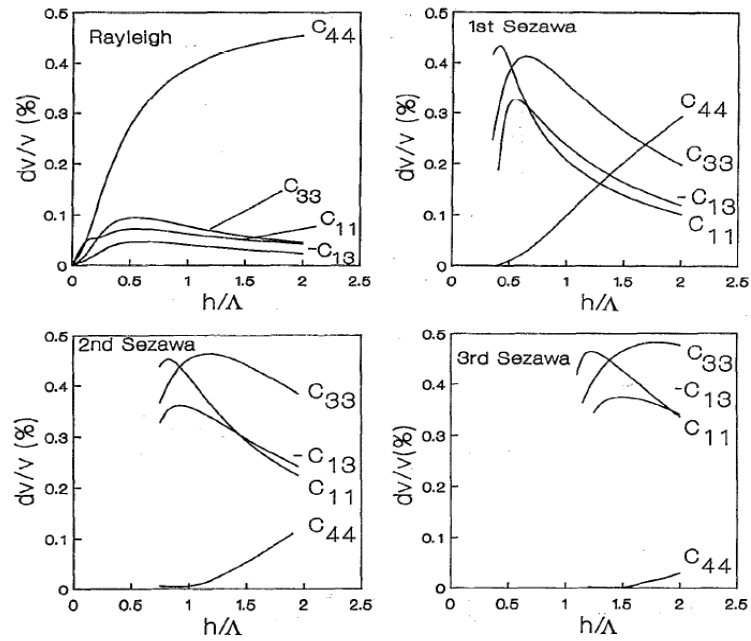

**Figure 10.** Relative changes of the phase velocity of the Rayleigh and first three Sezawa modes produced by a 1% increase of each elastic constant vs. the ratio between film thickness h and phonon wavelength. Calculations refer to an Ag/Ni superlattice film supported by a glass substrate. Reprinted with permission from [62], AIP Publishing, 1992.

A relatively easier situation may occur if the total thickness of the superlattice film becomes much larger than the light wavelength, so that the number of Sezawa modes is very large and the spectrum becomes almost continuum, as it would happen at the surface of a semi-infinite medium. In this case, as illustrated in Figure 11, the projection of the vertical acoustic displacement at the free surface contains the discrete peak of the Rayleigh surface wave, followed by a continuum of modes. Interestingly, in this continuum spectrum one can identify a dip, labeled Longitudinal Threshold (LT), occurring in correspondence of the frequency of the longitudinal acoustic wave propagating parallel to the free surface (i.e., where one would have observed the LM peak in a transparent film). The experimental counterpart of this calculation is in Figure 12, where the spectrum relative to a 5 μm thick AgNi superlattice film is shown [66]. It is clear that, in addition to the RW peak, one can see the dip corresponding to the LT and directly derive unambiguously the value of the longitudinal elastic constant $C_{11} = \rho V_{LT}^2$.

The same methodology, based on the detection of the discrete RW peak and the dip associated to the LT has been also used to selectively obtain the elastic constants $C_{11}$ and $C_{44}$ in polycrystalline Ta/Al superlattices with hexagonal elastic symmetry [67].

Finally, also for opaque semiconductors with a cubic elastic symmetry, such as epitaxial InGaAs films grown on GaAs analyzed in Ref. [68], the elastic constants have been obtained looking at both the RW and the LT. In addition, also the edge of the continuum shoulder (transverse threshold, TT) that is put in evidence in Figure 13, was measured. More recently, it has been shown that, taking advantage of the RW peak, the TT edge and the dip of the LT, a simple and robust fitting procedure has been presented for determining the three independent elastic constants of a cubic thick film from surf-BLS, using InAs0.91Sb0.09 as the case study [69].

To summarize the discussion of this Section about the potentiality of the technique, we provide in Table 1 a schematic presentation of the experimental methodology that is suitable to gain information about the elastic constants of thin films, considering their thickness and their transparency. However, before concluding this section, we want also to put in evidence a few weak aspects of surf-BLS. The first one is the limited accuracy in the determination of the phase velocities and of the phonon lifetime (attenuation), which are derived from the frequency position and the line width of the Brillouin peaks in the spectra, respectively. The accuracy in the determination of these quantities is limited to be above about 0.5% in the best cases, due to the intrinsic limitations in the frequency and wavevector resolution. It follows that the derived elastic constants are affected by uncertainty larger that about 1%. Another relatively weak point is the rather long acquisition times of each spectrum, that may range from a few tens of minutes to several hours (in the worst case of shear horizontal (Love) modes). Finally, the unavoidable heating of the sample region by the focused laser light should be taken into account. In fact, typical incident powers are between 100 and 300 mW for conventional measurements and between 1 and 10 mW for micro-focused ones. This causes a temperature increase ranging from a few degrees to a few tens of degrees, depending on the specific incident power and on the dissipation characteristics of the specimen.

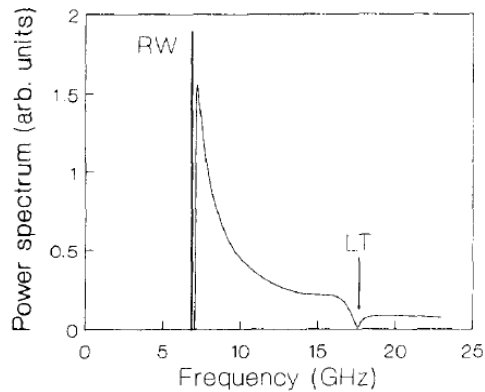

**Figure 11.** Calculated vertical component of the power spectrum of surface phonons, for a semi-infinite hexagonal medium (the phonon wavevector is fixed at 0.0226 nm$^{-1}$). The vertical line represents the position of the Rayleigh wave (RW); the arrow indicates the dip corresponding to the longitudinal threshold (LT). Reprinted with permission from [66], Elsevier, 1992.

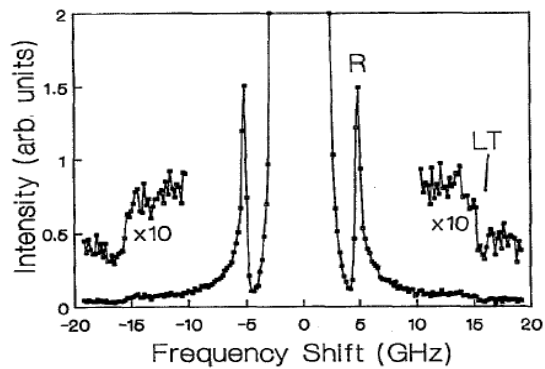

**Figure 12.** Brillouin spectrum from an unsupported Ag/Ni superlattice, 5 μm thick. Both the Rayleigh wave peak (R) and the dip corresponding to the longitudinal threshold (LT) are observable; the incidence angle is 67.5″, which corresponds to a value of h/Λ = 18. Reprinted with permission from [62], AIP Publishing, 1992.

**Table 1.** Methods of determination of the elastic constants for films of different symmetry, thickness and transparency. "Thin" and "thick" refer to films whose thickness is below about 500–600 nm and above about 1 micron. RW: Rayleigh wave; LM: longitudinal mode; LB: longitudinal bulk; LT: Longitudinal Threshold.

| Symmetry | Isotropic | Cubic | Hexagonal |
|---|---|---|---|
| Independent constants | $C_{11}$ $C_{44}$ | $C_{11}$ $C_{12}$ $C_{44}$ | $C_{11}$ $C_{13}$ $C_{22}$ $C_{44}$ $C_{66}$ |
| Thin transparent film * | $C_{11}$ and $C_{44}$ from the dispersion of the RW and SW | $C_{11}$ $C_{12}$ and $C_{44}$ from the dispersion of RW and SW | $C_{11}$ $C_{13}$ $C_{33}$ and $C_{44}$ from the dispersion of RW and SW |

| Thick transparent film ** | $C_{11}$ from the LM and/or LB $C_{44}$ from the RW | $C_{11}$ from the LM, $C_{44}$ from the RW, $C_{12}$ from the LB | $C_{11}$ from the LM, $C_{44}$ from the RW, $C_{13}$ and $C_{33}$ from the LB |
| --- | --- | --- | --- |
| Thin opaque film * | $C_{11}$ and $C_{44}$ from the dispersion of the RW and SW | $C_{11}$ $C_{12}$ and $C_{44}$ from the dispersion of RW and SW | $C_{11}$ $C_{13}$ $C_{33}$ and $C_{44}$ from the dispersion of RW and SW |
| Thick opaque film ** | $C_{44}$ from the RW and $C_{11}$ from the LT | $C_{11}$ from the LT, $C_{12}$ and $C_{44}$ from the RW and TT | $C_{44}$ from the RW and $C_{11}$ from the LT |

\* It is necessary to know the thickness and the mass density of the film; ** It is necessary to know the mass density of the film. Moreover, in order to exploit the information coming from the LB mode in transparent films, it is also necessary to know the refractive index.

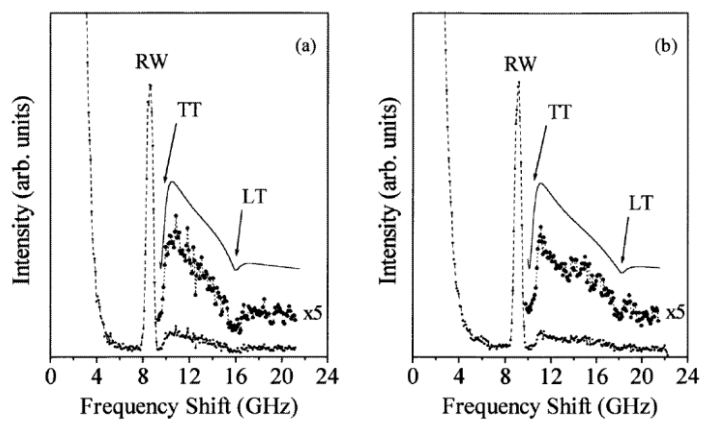

**Figure 13.** Experimental Brillouin spectra (dots) taken from a InGaAs/InP superlattice with periodicity 3.1 nm and total thickness 201.5 nm with Qs along [100] (**a**) and along [110] (**b**), for an angle of incidence θ = 67.5°. The solid lines correspond to the calculated perpendicular components of the surface phonon power spectrum, convoluted with a Gaussian function of width 0.6 GHz (the instrumental resolution). The Rayleigh wave peak (RW) the dip corresponding to the longitudinal threshold (LT) and the shoulder starting at the transverse threshold (TT) are indicated. Reprinted with permission from [68], IOP Publishing, 1996.

*3.4. Multilayered Structures*

Coming back to the complex layered structure of current devices such as acoustic resonators (Figure 1), one way to achieve a complete knowledge of the elastic properties of the stack is to prepare and analyze a series of single films with the different materials, studying each of them independently according to the procedure described in the previous paragraphs. Alternatively, one can even directly measure the entire stack in a surf-BLS experiment and compare the frequencies of the detected modes to those calculated solving acoustic propagation in the multilayer by a proper numerical approach. The case of a bilayer on a substrate was successfully analyzed, for instance, in Refs. [39,70]. The difficulty that may arise is that in this case the number of fitting parameters is so large (a few elastic constants for each constitutive layer) that, likely, it will not be possible to achieve a meaningful set of best fit parameters.

## 4. New Perspectives from Micro-Focused Surf-BLS: Mapping the Spatial Profile of Acoustic Modes

While the conventional surf-BLS technique, presented above, offers the possibility to measure the frequency of acoustic excitations with a well-defined wavenumber in a large frequency range (1–500 GHz), it has a poor lateral resolution, determined by the diameter of the laser spot, that is usually about 30–40 μm. Moreover, the signal is proportional to the intensity of the acoustic field, but it has no phase sensitivity. This means that in the case of inhomogeneous samples, such as artificially defined phononic crystals (PC) [71] or acoustic resonators [1–4], one cannot achieve the imaging of the spatial profile of the acoustic field. This limitation can be overcome using the recently developed micro-focused surf-BLS technique, whose schematic apparatus is drawn in Figure 14, that has been extensively used in the last decade to investigate the lateral localization of spin waves in ferromagnetic films and artificial magnonic crystals [72]. In fact, using a microscope objective with large numerical aperture, the laser light can be focused down to a diffraction-limited spot with a diameter below 300 nm and a submicrometric depth of focus. In addition, phase resolution can be achieved in the case of externally driven excitations, paving the way to the tridimensional mapping of magnetic or acoustic excitations by using surf-BLS as a scanning probe technique, in conjunction with an automated scanning stage for the sample. A coaxial viewing system based on a collimated LED light source (455 nm wavelength), a beam expander, and a CCD camera is used to obtain a direct visualization of the laser spot and of the sample region under investigation. Let's notice that, although micro-focusing permits to achieve spatial and phase resolution, the wavevector sensitivity is lost, because of both the small dimension of the focused spot and the very large collection angle. Therefore, micro-focused surf-BLS is well suited to analyze the acoustic field driven by an external signal (as in acoustic resonators), rather than to detect thermally activated phonons. The application to these aim is initiated and preliminary results have been just presented [73]. Further progress will be achieved in the near future, also because the operating frequencies of devices are expected to increase above a few GHz, so the availability of a technique operating in the frequency domain (rather than in the time domain as done up to now [74,75]) can be advantageous.

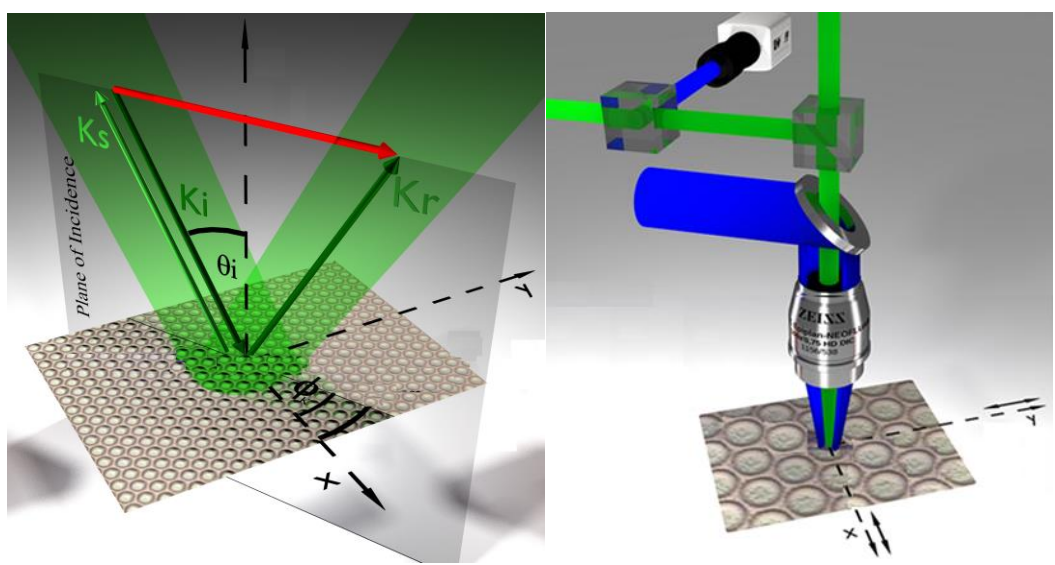

**Figure 14.** (Left) Scattering geometry of the conventional surf-BLS experiment. The wave vectors of the incident, reflected and scattered light are indicated, together with that of the acoustic surface phonon (in red). The illuminated area is a circle with a diameter of tens of microns (i.e., much larger than the wavelength of the detected phonons) so the wavevector conservation is guaranteed; (Right) Schematic drawing of the micro-focused apparatus. The light beams coming from either the laser or the auxiliary LED sources are represented by green and blue beams, respectively. The illuminated area is a spot of about 300 nm diameter and the collection angle is very large, so the conservation of momentum does not apply.

## 5. Conclusions and Future Perspectives

An overview on surf-BLS investigation of the elastic properties of single- and multi-layered materials has been presented, with emphasis given to the influence of the thickness and the transparency of the investigated films dependence on the characteristics of the spectra. Depending on these characteristics, different experimental strategies have been outlined, indicating how to get maximum information with minimum effort. In specific cases, it is possible to achieve a complete elastic characterization of the investigated samples: in other cases only a few elastic constants can be determined while others may remain out of the reach of this technique. Moreover, there has been a detailed mapping of the tridimensional spatial profile of the acoustic field in surface- or bulk-acoustic-wave resonators by the recently developed micro-focused scanning apparatus. The above capabilities could significantly contribute to the desired elastic characterization of multilayered acoustic structures operating at microwave frequencies in ubiquitous ICT devices, such as mobile phones. The reason why the exploitation of surf-BLS is still underestimated by industrial manufacturers in this field is, in my opinion, because it is not a standard characterization technique and requires a rather specific expertise. Therefore, I hope that this review paper may be of help especially for new researchers who want to take advantage of the potentialities of surf-BLS and establish a tight connection with industrial partners to contribute to the design and optimization of the next generation of acoustic resonators for the current and forthcoming generation of mobile devices.

**Conflicts of Interest:** The author declares no conflict of interest.